\documentclass{aa}  
\usepackage{graphicx}
\usepackage{epsfig}
\usepackage{txfonts}
\begin{document}

\title{First tentative detection of
anisotropy in the QSO distribution around nearby edge-on 
spiral galaxies}
   \subtitle{}
   \author{M. L\'opez-Corredoira,\inst{1} C. M. Guti\'errez\inst{1}}
   \offprints{martinlc@iac.es}

\institute{
$^1$ Instituto de Astrof\'\i sica de Canarias, C/.V\'\i a L\'actea, s/n,
E-38200 La Laguna (S/C de Tenerife), Spain}

   \date{Received xxxx; accepted xxxx}

  \abstract
   {}
   {To check whether the polar angle distribution of QSOs around nearby spiral 
   galaxies is isotropic or not.}
   {A statistical analysis of the polar angle distribution of large samples 
   of QSOs from the SDSS survey 
   and Monte Carlo simulations to calculate their significance are carried out.}
   {There is a clear excess of QSOs near the minor axis with respect to
the major axis of nearby edge-on spiral galaxies,
significant at a level 3.5$\sigma $ up to angular distances of $\sim 3^\circ$
(or $\sim 1.7$ Mpc) from the centre of each galaxy. 
The significance is increased to 3.9$\sigma $ with the $z>0.5$ QSOs, 
and it reaches 4.8$\sigma $ if we include galaxies whose circles of radius
3 degrees are covered by the SDSS in more than 98\% 
(instead of 100\%) of the area.}
   {Gravitational lensing in the halo of nearby galaxies or extinction 
   seem insufficient to explain the observed anisotropic distribution of 
   QSOs. The anisotropic distribution agrees qualitatively with the predictions of Arp's
   models, which claim that QSOs are ejected by galaxies along the rotation
   axis, although Arp's prediction give a distance of the QSOs $\sim 3$ times 
   smaller than that found here. 
   In any case, a chance fluctuation, although highly
   improbable, might be a possibility rather than a true anisotropy,
   and the present results 
   should be corroborated by other groups and samples, so we prefer to
   consider it as just  a first tentative detection.}
   
\titlerunning{Anisotropy/QSOs}
\authorrunning{L\'opez-Corredoira \& Guti\'errez}
   \maketitle
%

\section{Introduction}

The first hints of a possible relationship between
nearby galaxies and high redshift QSOs came from Arp (1966, 1967),
who observed radio sources across active galaxies that were 
identified as QSOs. 
Since the pioneering work by Burbidge et al.\ (1971), 
several groups have demonstrated the existence of an angular correlation 
between samples of QSOs and low redshift galaxies (e.g.\ the
reviews of Guimaraes 2005 and Burbidge 2001 respectively of orthodox
and heterodox approaches). 
This correlation extends  to angular 
scales of $\sim 1$ degree. Although weak
gravitational lensing by dark matter has been proposed to be the cause of these
correlations, many authors have found this hypothesis insufficient to
explain the correlations (Ben\'\i tez et al.\ 2001; Gazta\~naga 2003;
Nollenberg \& Williams 2005). The recent analysis carried 
out by Scranton et al.\ (2005), who
used Sloan Digital Sky Survey (SDSS) photometric 
data containing $\sim 2\times 10^5$ quasars and $\sim
1.3\times 10^6$ galaxies, showed that the amplitude and sign
of the angular correlation function between both sets of objects depend on the
magnitude limit considered for the sample of QSOs. 
From this correlation, Scranton et al.\ (2005)
proposed an  ad hoc  halo distribution function compatible with a
cross-correlation of very small amplitude ($\omega_{\rm GQ}<0.04$) 
of faint galaxies with QSO candidates
selected photometrically (5\% of this sample are not QSOs; Richards et al.\
2004). It is small because the mean separation among galaxies is small
and any positive correlation of QSOs around a galaxy is diluted with
the contamination of many other QSOs belonging to other galaxies.
This still seems insufficient to solve 
the most important problem of the correlation
found between QSOs and the {\it nearest and brightest} 
galaxies (Kovner 1989); for instance, for the high amplitude
angular correlation found by Chu et al.\ (1984) $\omega _{\rm GQ}\sim 5$.
Although  incompleteness could be responsible
for these correlations, it is a matter that requires further study.

The existence of a correlation between samples of objects with different redshifts
has been advocated by the supporters of non-cosmological redshifts 
(e.g.\ Burbidge 1999) as clear evidence of physical association and as proof that
QSOs are being ejected by low redshift galaxies. One variant of the model
(e.g.\ Arp 1998a, ch. 3; Arp 1999a) assumes that QSOs are ejected 
along the rotation axis of the parent galaxy, decreasing in redshift as 
they move outwards and eventually becoming normal galaxies.

There have been claims of anisotropy in the distribution of QSOs around
nearby galaxies. For instance, there are configurations of QSOs aligned along the
minor axis of a central Seyfert whose probability of being accidental is $10^{-9}$ to $10^{-10}$ 
(Arp 1998b, 1999a). Arp \& Hazard (1980) have reported three QSOs 
in a straight line, together with another set of three QSOs  in a straight line, a
configuration that is very difficult to explain as random. Arp (1999b)
finds that the probability of having six out of six QSOs aligned within $\pm 15^\circ $ of
the minor axis of NGC 5985 to be only $10^{-8}$ to $10^{-9}$. Arp \& Russell
(2001) found that the bright radio quasars 3C37 and 3C39 are paired across the
centroid of the disturbed galaxy pair NGC 470/474 in an arrangement with a
probability of only $2\times 10^{-9}$ of being accidental. There is anisotropy in
the radio QSO distribution at high flux densities (Shastri \& Gopal 1983): the
number of QSOs on one side of the M33 region is far greater ($\sim 11\sigma $) than
that of the diametrically opposite region. 
This alignment of QSOs with the parent galaxy would involve 
conservation of momentum in their ejection in the heterodox Arp/Burbidge/Narlikar
scenario (Narlikar \& Das 1980).

In this paper we undertake the first study of the polar angle distribution of 
QSOs around bright  nearby galaxies in statistical terms using large samples of both types of
objects. With the release of the SDSS survey, a systematic search for this effect 
is now possible. The motivation of this research is to test one
of the predictions given by the above authors concerning the ejection model.
The predictions are as follows: if the QSOs were
background objects, no preferential orientation in principle should be found;
however, if the QSOs were ejected by the galaxies, as Arp 
proposed (or alternatively a gravitational effect were predominant in the 
rotation axis of a galaxy), a higher concentration of objects near the 
rotational axis of the galaxy would be expected.

Section~2 presents the data and
sample selection; the analysis and statistical tests are presented in Sections~3
and 4, Section~5 discusses some details of the method
to measure the anisotropy, and in Section 6 we interpret the results.

\section{Data and sample selection}

Two sources of public data have been used in this work: 
\begin{itemize}
\item {\bf Third Reference Catalog of Bright Galaxies (de Vaucouleurs et al.\ 1991,
updated on 16 February 1995; hereafter RC3)}: this catalogue is complete for galaxies with
apparent diameters greater than  1 arcmin at the $D_{25}$ isophotal level and total $B$-band
magnitudes $B_T\le 15.5$, with a redshift $z\le 0.05$. It contains a list of 23\,011 galaxies over the whole sky.

\item {\bf Spectroscopic catalogue of the Third data release (DR3)  of the Sloan
Digital Sky Survey (Abazajian et al.\ 2005; hereafter SDSS):} this catalogue   covers
4188 square degrees and contains 51\,027 QSOs. The spectra have  signal-to-noise ratios
$SNR>4$ per pixel at $m_g=20.2$. From these QSOs, we took the vetted subsample
by Schneider et al.\ (2005) in which only those objects with $M_i<-22$ are selected 
(this implies that the low redshift QSOs are removed, and that the redshift range is 
from 0.08 to 5.41;
their redshift is therefore always higher than those of the RC3 galaxies) and many of
them have been checked manually to be QSOs (or otherwise removed). The total
number of QSOs of this vetted catalogue is 46\,420. We prefer to use the
spectroscopic data instead of the photometric data used, for example, by Scranton et al.\
(2005) with a colour selection technique to separate the QSOs for reliable
identification and an accurate estimate of the redshift.  While we were working with
these data, a fourth release of the survey was delivered that covered an area 14\%
larger (Adelman-McCarthy et al.\ 2006), but an updated  vetted subsample has not been
produced yet (and will not be produced until a fifth release of the data; D. P.
Schneider, priv.\ comm.).  In \S \ref{.tests} we show some
results of the unvetted 4th release.
\end{itemize}  

We are aware that the DR3 Quasar spectroscopic catalogue is not a statistical sample.
Its QSOs do not all follow the same target selection, and the completeness
depends on the positions, magnitudes, redshifts, etc., of the QSOs. Nonetheless, it is
appropriate for the statistical analysis presented here because that completeness is related
to the intrinsic properties of the QSOs and not to the  orientation of nearby galaxies.
Possible bias due to incompleteness and other systematic effects  in the sample will be
considered and quantified through Monte Carlo simulations in Section \ref{.tests}.  

We take the RC3 catalogue and select all spiral galaxies with information on
radial velocity ($v_{\rm 3K}$), the angular radius of the 25th mag isophote ($\theta
_{25}$), the ratio between the major and minor axes and position angle, and whose
coordinates are such that a circle of radius equal to $\theta _{\rm max}$ around  each galaxy
was totally covered by the 3rd release of the SDSS survey (see \S \ref{.incomp} for
the use of circles partially covered). Spiral galaxies were
chosen  because the inclination of their discs is easily derived from the projected
axial  ratio ($\cos i=\frac{b}{a}$). We reject the very nearby
galaxies  with $v_{\rm 3K}<700$ km/s (because their distances cannot be determined
from their redshifts and because they would also extend too much in angular size with 
respect to the average sample; in any case, they are very few in number, and their inclusion
would not significantly alter the statistics in this paper). 
We also select edge-on galaxies within the range $\pm 25^\circ $ ($\sin i>0.906$).
There is no convenient  a posteriori selection of
galaxies: the selection of edge-on galaxies is necessary because they allow a clear
separation between the rotation axis and the disc of the galaxy (the projection of
the rotation axis is perpendicular to the disc of the galaxy and has a negligible
component in the line of sight), and only spiral galaxies provide 
information for deriving the inclination to make possible the separation of edge-on galaxies.
A range of
25 degrees ($i>65$ deg) was taken in order to achieve a high enough 
number of galaxies within the
nearly edge-on criterion ($\sin i>0.906$). If the limiting inclination were
slightly larger or  smaller, the statistical results
given in this paper would be similar. If we selected only galaxies with an
inclination very close to 90 degrees, we would have very few galaxies 
in our sample and the 
statistics would be much poorer; if we took the limiting inclination much 
lower than 65 degrees, we would introduce noise from nearly face-on
galaxies.

\begin{figure} 
\begin{center} 
\vspace{1cm} 
\mbox{\epsfig{file=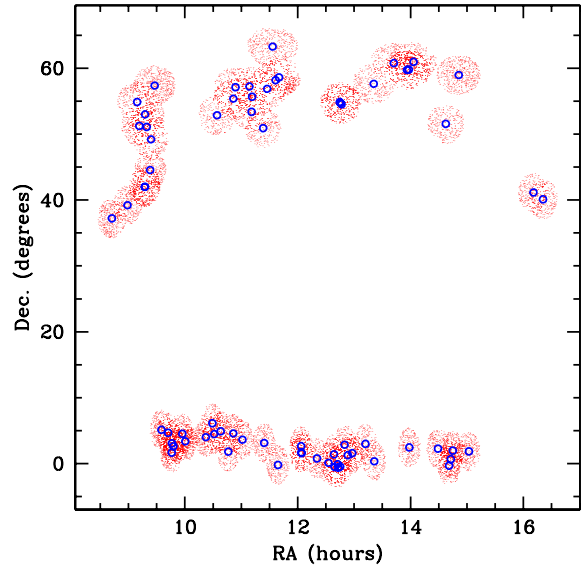,height=8cm}}
\end{center} 
\caption{Position on the sky of the sample of objects analysed in this article. Large
open circles represent galaxies, and small dots QSOs. The plot corresponds to the
case $\theta_{\rm max}=3^\circ$.} 
\label{sample}
\end{figure}

We select the 44\,975 QSOs with $m_g\le 21$  within an angle of  $\theta
_{25}<\theta<\theta_{\rm max}$  from each galaxy (the lower limit in distance  avoids possible effects of
galactic extinction in background QSOs). There is no overlap in the redshift of the
galaxies ($z<0.05$) and QSOs ($z>\approx 0.08$). Figure~\ref{sample} shows the
position of the resulting samples for the case $\theta_{\rm max}=3^\circ$.

\section{Analysis}
\label{.indiv}

We compute the position angle of QSOs with respect to the minor axis of each galaxy.
For $\theta _{\rm max}=3^\circ$ the total number of QSO--galaxy pairs is 25\,176 
for 71 RC3-galaxies. These 71 galaxies follow these
constraints, with an average distance of 46 Mpc (derived from $v_{\rm 3K}$, assuming a
Hubble constant of 72 km s$^{-1}$  Mpc$^{-1}$) and a median distance of 32 Mpc.
Some QSOs are counted more than once because  they are within the radius $\theta_{\rm max}$ 
of two or more galaxies (see Fig.~\ref{sample}), 
but their relative position angles with respect to
the different galaxies with which they are associated are uncorrelated, 
so they count as independent measures (see discussion in \S \ref{.twice}).

\begin{figure}
\begin{center}
\vspace{1cm}
\mbox{\epsfig{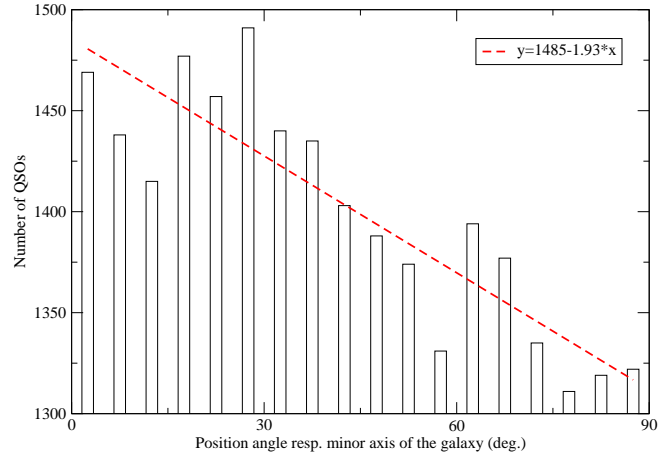}}
\end{center}
\caption{Histogram representing the number of QSOs with $m_g<21$
within a circle of radius equal to 3 degrees of 71 RC3 galaxies 
as a function of position angle with respect to the minor axes of 
these galaxies.}
\label{Fig:histoPA}
\end{figure}

A histogram of all counts vs.\ relative position angle with respect to the minor
axis of the galaxy is shown in Fig. \ref{Fig:histoPA} for  $\theta
_{\rm max}=3^\circ$.  Clearly, there is a decrease in counts as we move from the minor
to the major axis. The excess of counts in the minor axis direction is
$\approx 13$\% higher than in the major axis direction. A  fit for the counts per bin vs.\ position angle (PA)
with a linear function 

\begin{equation}
C=C_0(1+\alpha r)
\label{C}
,\end{equation}
\[
{\rm where}~r=\left(1-\frac{PA}{90^\circ }\right)\ {\rm (between\ 0\ and\ 1)}
\]
gives $C_0=1311\pm 13$, $\alpha =0.132\pm 0.017$. This means a detection of
anisotropy at the 7.8$\sigma $ level, although this is only the 
statistical significance of the survey, which does not represent
the statistical significance of the real (unbiased) distribution
of QSOs, as discussed in Section
\ref{.tests}.

\begin{figure}
\begin{center}
\vspace{1cm}
\mbox{\epsfig{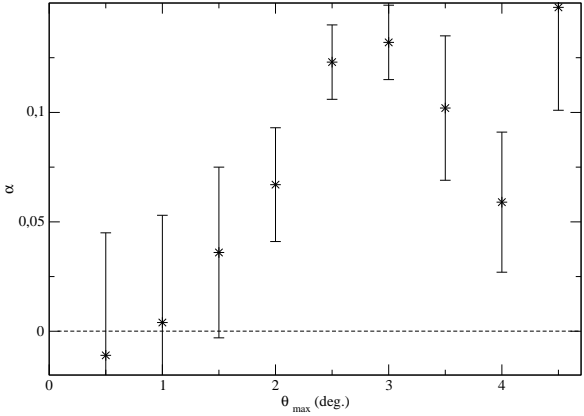}}
\end{center}
\caption{$\alpha $ vs.\ $\theta _{\rm max}$ for QSOs with $m_{g}\le 21$ 
and any redshift (the points are not independent).}
\label{Fig:alfa_th}
\end{figure}

Other values of $\theta _{\rm max}$ (see Fig.~\ref{Fig:alfa_th}) show the existence of anisotropy  but the best
significance is for 
$\theta _{\rm max}\approx 3^\circ$. For lower angles, there seems to be a lower value
of $\alpha $, although for $\theta _{\rm max}< 1^\circ$ the number of QSOs is too low
to draw any conclusions. For $\theta _{\rm max}>
4^\circ$ the number of galaxies with
total SDSS covered area in the circle with radius $\theta _{\rm max}$ is much lower
(for $\theta _{\rm max}=4.5^\circ $ we have only 12 galaxies). In any case, the
detection of $\alpha \ne 0$ (anisotropy) is clear, and the projected linear scale of
around 1--1.5 Mpc  (around 1.5--2.0 Mpc without projection) seems to have the
highest ratio of anisotropy.

\begin{figure}
\begin{center}
\vspace{1cm}
\mbox{\epsfig{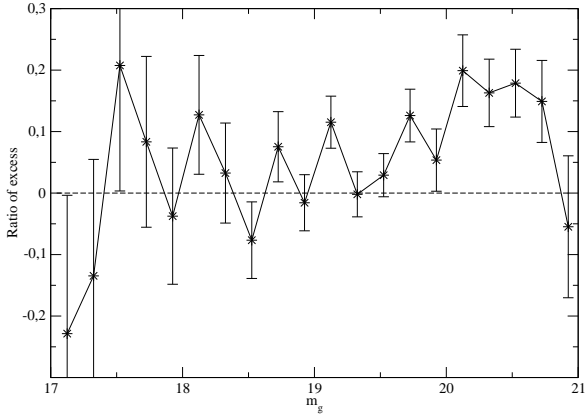}}
\end{center}
\caption{Ratio of relative excess of QSOs in the range of position angles
with respect to the minor axis 0--45$^\circ $
over the range 45--90$^\circ $
versus magnitude of QSOs.}
\label{Fig:ratio_mg}
\end{figure}

\begin{figure}
\begin{center}
\vspace{1cm}
\mbox{\epsfig{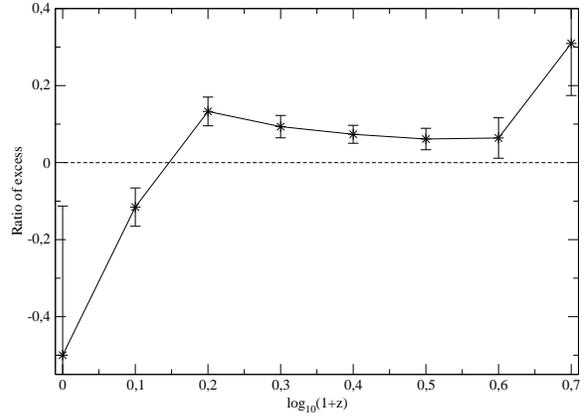}}
\end{center}
\caption{Ratio of relative excess of QSOs in the range of position angles
with respect to the minor axis 0--45$^\circ $
over the range 45--90$^\circ $ versus redshift of the QSOs.}
\label{Fig:ratio_z}
\end{figure}

The excess in the range 0--45$^\circ $ (13\,025 cases) over the range 45--90$^\circ $
(12\,151 cases) in Fig. \ref{Fig:histoPA} is $E_{45}=+874$ cases, a ratio of excess
of 7\%. These extra 874 sources are preferably placed  in the magnitude range 
$m_g>19.4$ and the range  $0.2<log_{10}(1+z)<0.5$ (i.e.\ $0.6<z<2.2$),
as can be observed in Figures \ref{Fig:ratio_mg}
and \ref{Fig:ratio_z}. 
Curiously, the QSOs with $z<\approx
0.5$ significantly  show the opposite trend on average: the excess is towards the
major axis. The peaks with significant excess (higher signal/noise)  or
relative excess are more or less for $m_g\sim 20.1$ and $z\sim 0.6$ respectively.

\begin{figure}
\begin{center}
\vspace{1cm}
\mbox{\epsfig{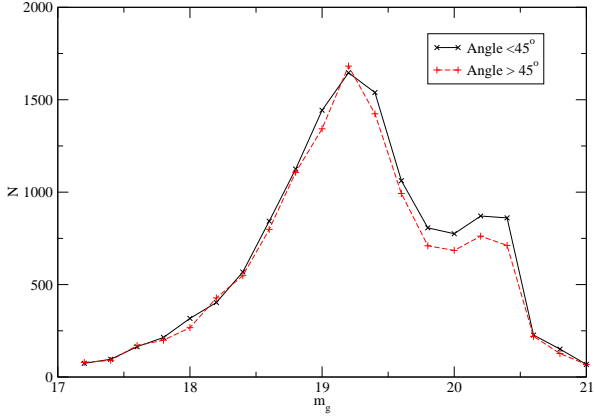}}
\end{center}
\caption{Differential QSO counts in bins of 0.2 mag around the
71 galaxies that follow $\theta _{\rm max}=3^\circ $. The angle
is with respect to the minor axis.}
\label{Fig:counts}
\end{figure}

\begin{figure}
\begin{center}
\vspace{1cm}
\mbox{\epsfig{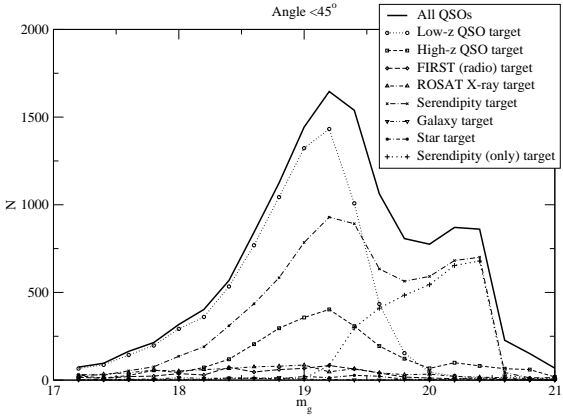}}
\end{center}
\caption{Differential QSO counts in bins of 0.2 mag  around the
71 galaxies which follow $\theta _{\rm max}=3^\circ $ with an angle
respect to the minor axis of less than 45 degrees. Different SDSS samples
 attending to different criteria to preselect the targets 
are used.}
\label{Fig:counts-target}
\end{figure}

Figure \ref{Fig:counts} also shows how the QSOs with anisotropy are preferentially
those with $m_g>19.4$. The limit is near the maximum of the differential
QSO counts; over 19.2 the counts are lower because of appreciable incompleteness (otherwise
the differential counts should increase monotonically). One might suspect that
the anisotropy has something to do with the incompleteness/bias of QSOs; however, it
will be demonstrated in \S \ref{.testangle} that the main cause of the anisotropy is not that;
at least the effects of incompleteness/bias, if there are any, are not enough to explain the
observed anisotropy.
Examining Fig. \ref{Fig:counts}, one should also realize that there are two structures
in the counts: two overlapping peaks, one with  a maximum at $m_g\approx 19.2$ and another 
at $m_g\approx 20.2$. Apparently, it is the second group of QSOs that is responsible
for the anisotropy, and this is shown over $m_g>19.4$ because this is the range where
the number of QSOs in the second group is relatively significant.
We suggest this to be the main reason for the manifestation of the anisotropy over $m_g>19.4$
rather than the incompleteness. 
This second peak corresponds mainly to the criterion of target selection in 
the SDSS survey (Schneider et al.\ 2005) of serendipity sources
with unusual colours; other targets do not make a such an important contribution
(see Fig. \ref{Fig:counts-target}). Values of $\alpha $ with the different
target selections are given in Table \ref{Tab:constr}.
When we take the QSOs that appear in the Serendipity target and
exclude the other targets, we get a value of $\alpha =0.380\pm 0.055$,
an excess of 38\% towards the minor axis instead of the average of 13\% with
the whole sample.

If we do a new analysis of the anisotropy with the added constraint that $m_g>19.4$,
$0.2<log_{10}(1+z)<0.5$ (a total of 13\,308 cases), we get 7509 cases of
QSO--galaxy association, and  $\alpha $
to 0.313$\pm 0.042$. This represents more than a 30\% excess towards the minor axis with
respect to the major axis. The analysis of the anisotropy only with the further added constraint that
$m_g>19.4$ gives 10\,204 cases of QSO--galaxy association with
$\alpha =0.237\pm 0.036$. If we restrict the analysis with 
the added constraint that $z>0.5$ ($\log
(1+z)>0.176$)  we get 22\,953 cases of QSO--galaxy association and
 $\alpha =0.156\pm 0.017$. Although this simple
constraint (in which we remove only 9\% of the associations with respect to the
analysis of the full sample) does not give a much higher value of $\alpha $; it
is much more significant statistically, as we comment in Section \ref{.tests}.
The results of these and other analyses
with different $\theta_{\rm max}$ are summarized in Table \ref{Tab:constr}. 

\begin{table*}
\caption{Anisotropy of QSOs with different constraints}
\begin{center}
\begin{tabular}{ccccc}
\label{Tab:constr}
$\theta_{\rm max}$ & Other constraints  & Number of pairs & $\alpha$  \\
\hline
0.5$^{\circ}$  &  & 2807 & $-0.011\pm 0.056$\\
1.0$^{\circ}$  & & 8698 & $0.004\pm 0.049$\\
1.5$^{\circ}$  & & 14902 & $0.036\pm 0.039$\\
2.0$^{\circ}$  & & 20562 & $0.067\pm 0.026$\\
2.5$^{\circ}$  & & 23819 & $0.123\pm 0.017$\\
3$^{\circ}$    &                                & 25176 & $0.132\pm 0.017$ \\
3$^{\circ}$    &         $m_g>19.4$               & 10204 & $0.237\pm 0.036$ \\
3$^{\circ}$    & $m_g>19.4$,  $0.2<\log _{10}(1+z)<0.5$ &  7509 & $0.313\pm 0.042$ \\
3$^{\circ}$    & $z>0.5$                        & 22953 & $0.156\pm 0.017$ \\
3$^{\circ}$    & Low-z QSOs target            & 15434 & $0.064\pm 0.026$ \\
3$^{\circ}$    & High-z QSOs target            & 5055 & $0.087\pm 0.044$ \\
3$^{\circ}$    & FIRST (radio) target            & 1184 & $0.203\pm 0.085$ \\
3$^{\circ}$    & ROSAT (X-ray) target            & 1728 & $0.031\pm 0.087$ \\
3$^{\circ}$    & Serendipity target            & 14445 & $0.179\pm 0.030$ \\
3$^{\circ}$    & Serendipity (only) target        & 5859 & $0.380\pm 0.055$ \\
3$^{\circ}$    & Star target            & 304 & $0.164\pm 0.200$ \\
3$^{\circ}$    & Galaxy target            & 205 & $-0.303\pm 0.207$ \\
3.5$^{\circ}$  & & 21286 & $0.102\pm 0.033$\\
4$^{\circ}$    & & 14130 & $0.059\pm 0.032$\\
4.5$^{\circ}$  & & 9328 & $0.148\pm 0.047$\\
\end{tabular}
\end{center}
\end{table*}

\begin{figure}
\begin{center}
\vspace{1cm}
\mbox{\epsfig{file=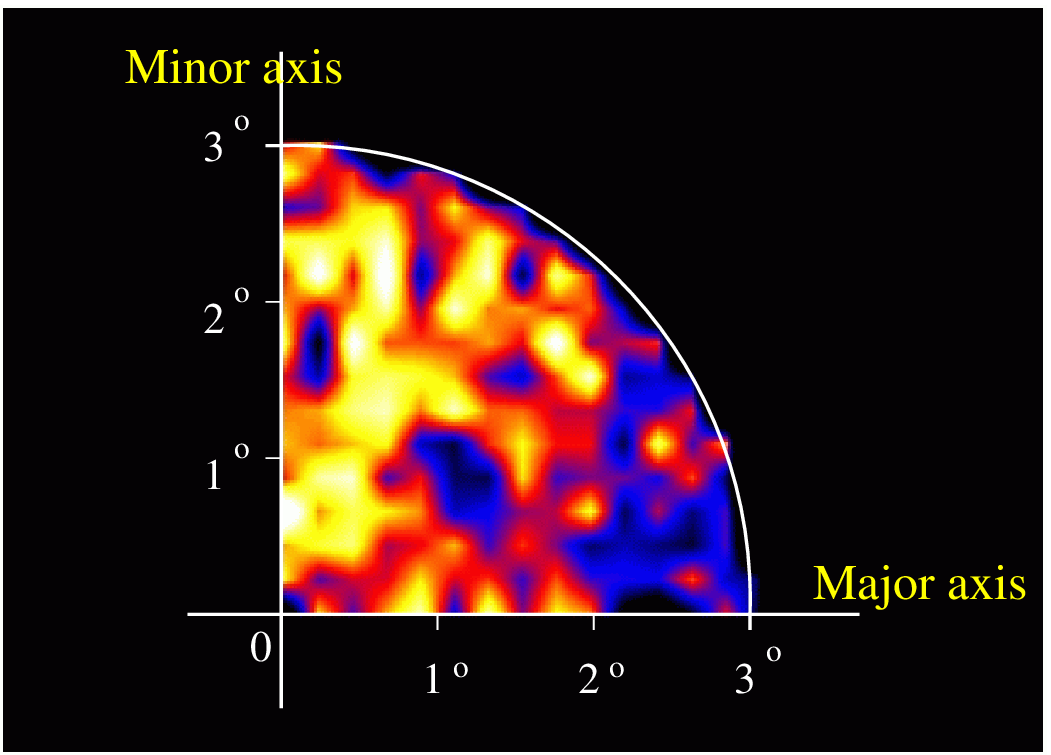,height=6cm}}
\mbox{\epsfig{file=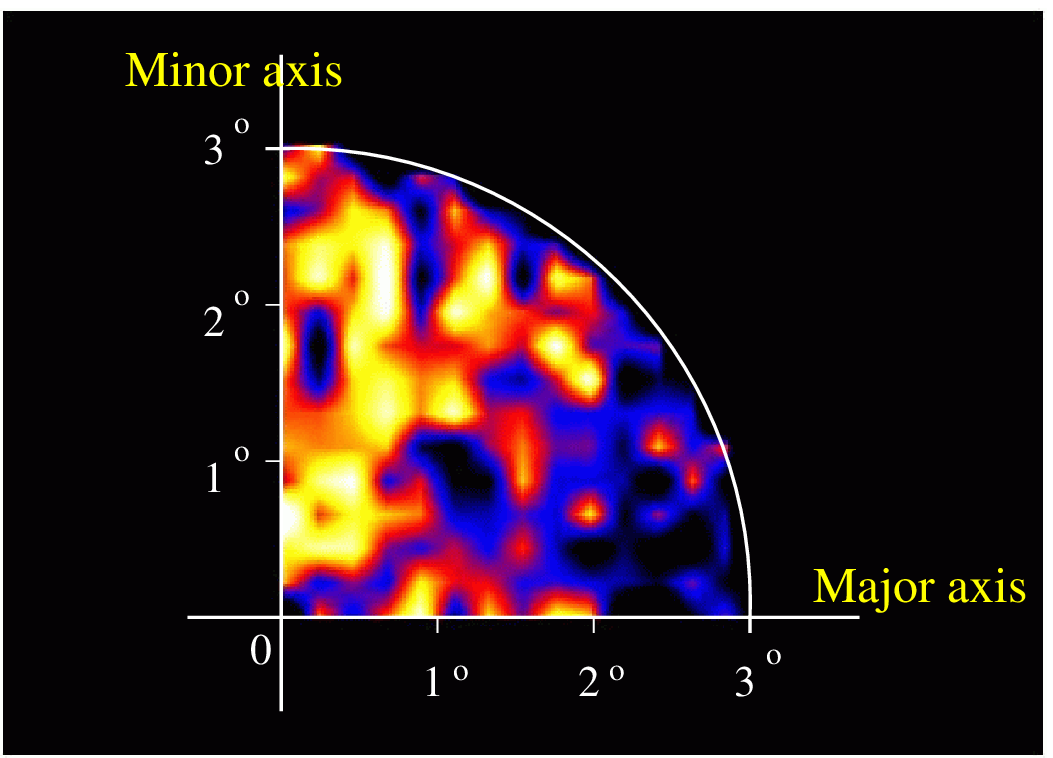,height=6cm}}
\end{center}
\caption{$Top:$ Counts of QSOs (total: 25\,176) as a function of position with
respect to the minor/major axis angles of the corresponding 71 galaxies 
(see text). Counts were plotted in bins of $0.2^\circ \times 0.2^\circ $
(with an average of around 140 QSOs per bin) and smoothed/interpolated.
The clearer colours indicate higher density. Note that
towards the minor axis the average density of QSOs is greater than towards the
major axis. Most of the fluctuations are presumably statistical (Poissonian).
$Bottom:$ The same but with the extra constraint $z>0.5$ (total: 22\,953 counts).}
\label{Fig:twod}
\end{figure}

The two-dimensional plot  of the polar and radial distribution
for two of the cases analysed are presented in Figure~\ref{Fig:twod}.
Apparently, QSOs are not randomly distributed around galaxies  but show a 
quite significant trend to follow the minor axes of those galaxies. Since the
median distance of the galaxies is 32 Mpc, the median linear projected maximum
distance corresponding to $\theta _{\rm max}=3^\circ$ is 1.7 Mpc. 

\begin{figure}
\begin{center}
\vspace{1cm}
\mbox{\epsfig{file=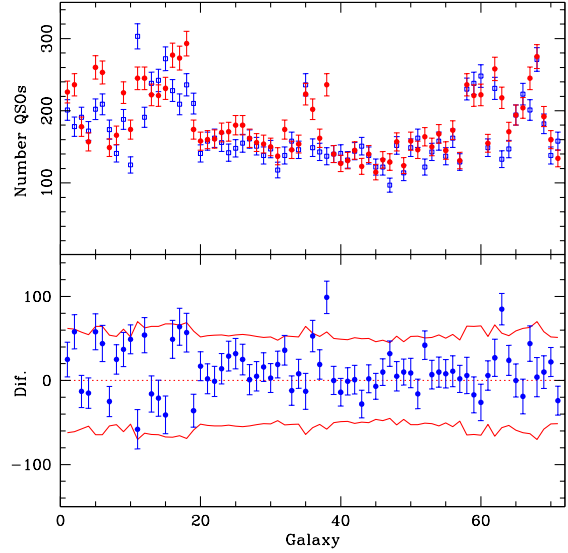,height=8cm}}
\end{center}
\caption{$Top:$ Number of QSOs at $\pm 45^\circ$ from the minor axis 
({\it open points}), and
from the major axis ({\it filled points}) around the minor and major axes 
for the 71 galaxies 
ordered in right ascension considered in the case $\theta_{\rm max}=3^\circ$. 
$Bottom:$ Difference between number of QSOS around the minor and major axes. 
The continuum lines represent the $\pm 3\sigma$ level.}
\label{fig_indiv_gal}
\end{figure}

Figure~\ref{fig_indiv_gal} presents for the 71 galaxies considered in the case
$\theta_{\rm max}=3^\circ$ the number of QSOs at $\pm 45^\circ$ from the minor and major 
axes respectively. The values
are also given in Table \ref{Tab:indiv}. 
There is in Fig.\ \ref{fig_indiv_gal} a different degree 
of dispersion of anisotropies depending on the number of galaxies (galaxies
with number less than 20 and higher than 60 present in general more
dispersion in anisotropy). This may be due to
the different levels of completeness for the different galaxies;
the galaxies are numbered in Fig.\ \ref{fig_indiv_gal} in right ascension 
order, so this most probably means that there are some regions of
the sky in which the completeness is lower, and the number of background
sources is also lower, as observed.

\begin{table}[htb]
\caption{Distribution of SDSS QSOs with $m_g\le 21$ within $\theta _{25}<\theta
<3^\circ $
around the 71 selected spiral galaxies (with 
$i>65^\circ $ that have complete coverage of circles of
radius $\theta _{\rm max}=3^\circ $). $E_{45}$: excess of QSOs
in the region which are at position angles less than 45$^\circ $ from
the minor axis over those that have position angles greater
than 45$^\circ $ from it (with Poissonian error). The galaxy
type or environment is taken from the SIMBAD database classification.
Distance ($d$) derived from the redshift with a Hubble constant
72 km/s/Mpc. PA is the position angle of the major axis of the
galaxy.}
\begin{center}
\begin{tabular}{ccccccc}
\label{Tab:indiv}
\# & PGC-\# & Type/env.& $d$ (Mpc) & PA($^\circ $) & QSOs & $E_{45}$  \\ \hline
1 & 24453 & -- & 54.6 & 72 & 427  & +25($\pm $21) \\
2 & 25232 & -- & 11.1 & 115 & 414  & +58($\pm $20) \\
3 & 25781 & -- & 35.8 & 120 & 369  & -13($\pm $19) \\
4 & 25910 & Low S.Br.  & 32.6 & 98 & 329 & -15($\pm $18) \\
5 & 26232 & Interact. & 27.0 & 160 & 462 & +58($\pm $21) \\
6 & 26238 & in Pair & 27.8 & 125  & 462 & +44($\pm $21) \\
7 & 26246 & in Pair & 34.4 & 48 & 323 & -25($\pm $18) \\
8 & 26351 & -- & 9.9 & 140  & 307 & +25($\pm $18) \\
9 & 26563 & -- & 40.1 & 4 & 413 & +37($\pm $20) \\
10 & 26631 & Interact. & 40.5 & 50 & 299 & +49($\pm $17) \\
11 & 26856 & -- & 46.6 & 123 & 548 & -58($\pm $23) \\
12 & 27248 & -- & 32.3 & 47  & 436 & +50($\pm $21) \\
13 & 27734 & in Pair & 32.8 & 72 & 460 & -16($\pm $21) \\
14 & 28010 & -- & 30.6 & 127 & 463 & -21($\pm $22) \\
15 & 28033 & in Group & 88.3 & 68 & 503 & -41($\pm $22) \\
16 & 28148 & -- & 30.7 & 15  & 505 & +49($\pm $22) \\
17 & 28741 & -- & 34.5 & 18  & 482 & +64($\pm $22) \\
18 & 28939 & -- & 33.1 & 151 & 529 & +57($\pm $23) \\
19 & 30364 & -- & 99.7 & 62 & 384 & -36($\pm $20) \\
20 & 30885 & -- & 54.3 & 111  & 299 & +17($\pm $17) \\ 
21 & 31037 & -- & 21.2 & 168 & 318 & -2($\pm $18) \\
22 & 31269 & -- & 101.3 & 156 & 323 & -1($\pm $18) \\  
23 & 31608 & -- & 96.6 & 125 & 326 & +14($\pm $18) \\
24 & 32153 & -- & 18.6 & 159 & 313 & +29($\pm $18) \\
25 & 32564 & -- & 42.2 & 78 & 328 & +32($\pm $18) \\
\end{tabular}
\end{center}
\end{table}

\begin{table}[htb]
\begin{center}
Cont. Table \protect{\ref{Tab:indiv}}.
\begin{tabular}{ccccccc}
\# & PGC-\# & Type/env.& d (Mpc) & PA($^\circ $) & QSOs & $E_{45}$  \\ \hline
26 & 32570 & -- & 19.3 & 130 & 335 & +25($\pm $18) \\
27 & 32714 & in Group & 28.8 & 48 & 323 & +1($\pm $18) \\
28 & 33234 & -- & 20.7 & 20 & 307 & +5($\pm $18) \\
29 & 33766 & -- & 29.2 & 99 & 292 & +16($\pm $17) \\
30 & 33964 & -- & 42.3 & 38 & 297 & +3($\pm $17) \\
31 & 34030 & -- & 12.0 & 80 & 255 & +19($\pm $16) \\
32 & 34971 & -- & 13.9 & 142 & 312 & +36($\pm $18) \\
33 & 35037 & -- & 154.5 & 144  & 304 & -12($\pm $17) \\
34 & 35249 & -- & 25.3 & 128 & 300 & +8($\pm $17) \\
35 & 35675 & -- & 19.5 & 60  & 459 & -13($\pm $21) \\
36 & 35900 & -- & 19.2 & 9 & 351 & +53($\pm $19) \\
37 & 36102 & -- & 80.4 & 132 & 305 & +19($\pm $17) \\
38 & 36192 & -- & 17.3 & 53 & 373 & +99($\pm $19) \\
39 & 38117 & in Cluster & 21.9 & 149  & 280 & 0($\pm $17) \\
40 & 38120 & in Cluster & 87.8 & 140 & 268 & -14($\pm $16) \\
41 & 38188 & in Cluster & 89.4 & 100 & 263 & -1($\pm $16) \\
42 & 39832 & Low S.Br. & 34.1 & 101  & 289 & +1($\pm $17) \\
43 & 41618 & in Pair & 20.3 & 83 & 274 & -28($\pm $17) \\
44 & 42255 & --  & 76.5 & 148 & 278 & +2($\pm $17) \\
45 & 42336 & -- & 19.6 & 97 & 237 & -7($\pm $15) \\
46 & 42689 & in Group & 28.5 & 63 & 254 & +10($\pm $16) \\
47 & 42791 & in Group & 41.4 & 37 & 226 & +32($\pm $15) \\
48 & 42942 & -- & 71.7 & 177 & 309 & +5($\pm $18) \\
49 & 42975 & LINER & 25.7 & 42 & 238 & +30($\pm $15) \\
50 & 42998 & --  & 69.0 & 97  & 307 & +9($\pm $18) \\ 
\end{tabular}
\end{center}
\end{table}

\begin{table}[htb]
\begin{center}
Cont. Table \protect{\ref{Tab:indiv}}.
\begin{tabular}{ccccccc}
\# & PGC-\# & Type/env.& d (Mpc) & PA($^\circ $) & QSOs & $E_{45}$  \\ \hline
51 & 43101 & -- & 71.9  & 3 & 308 & -16($\pm $18) \\
52 & 43397 & -- & 20.6  & 0 & 286 & +22($\pm $17) \\
53 & 43784 & -- & 20.3  & 133  & 293 & +7($\pm $17) \\
54 & 44392 & -- & 21.4  & 89 & 326 & +10($\pm $18) \\
55 & 45844 & -- & 46.5  & 32 & 282 & -13($\pm $17) \\
56 & 46589 & Low S.Br. & 31.4 & 153  & 335 & +8($\pm $18) \\
57 & 46633 & -- & 81.8 & 170 & 260 & +2($\pm $16) \\
58 & 48534 & -- & 32.1 & 30 & 466 & +6($\pm $22) \\
59 & 49548 & -- & 27.1 & 3  & 459 & -17($\pm $21) \\
60 & 49712 & -- & 43.4 & 167 & 470 & -26($\pm $22) \\
61 & 49758 & -- & 103.9 & 79 & 304 & +6($\pm $17) \\
62 & 50069 & -- & 30.3 & 174 & 489 & +27($\pm $22) \\
63 & 51752 & -- & 114.5 & 16 & 351 & +85($\pm $19) \\
64 & 52266 & in Pair & 32.3 & 35 & 318 & +24($\pm $18) \\
65 & 52455 & Low S.Br. & 27.3 & 107 & 388 & 0($\pm $19) \\
66 & 52558 & -- & 28.1 & 66 & 427 & -19($\pm $21) \\
67 & 52665 & in Pair & 27.0 & 170 & 446 & +44($\pm $21) \\
68 & 53043 & -- & 30.8 & 144 & 546 & +1($\pm $23) \\
69 & 53683 & -- & 20.0 & 115 & 374 & +10($\pm $19) \\
70 & 57386 & in Cluster  & 135.0 & 173 & 298 & +22($\pm $17) \\
71 & 57882 & in Cluster & 141.9 & 65 & 292 & -24($\pm $17) \\ \hline
All & ------ & ------ & --- & --- & 25176  & +874($\pm $159) \\ \hline
\end{tabular}
\end{center}
\end{table}

\begin{figure}
\begin{center}
\vspace{1cm}
\mbox{\epsfig{file=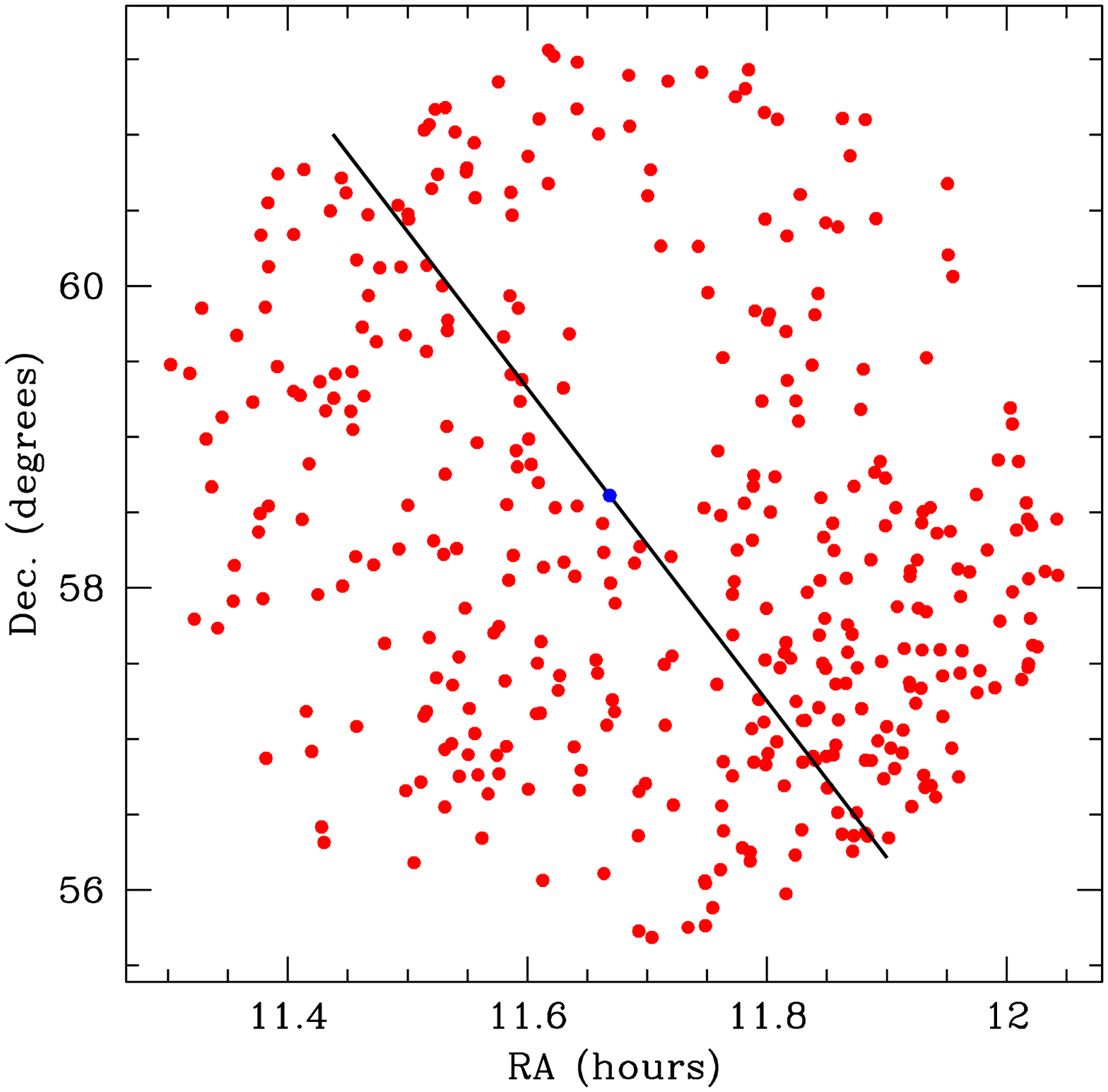,height=6cm}}

\vspace{1cm}

\mbox{\epsfig{file=histoPA_NGC3795.eps,height=6cm}}
\end{center}
\caption{Top: Distribution of QSOs around NGC 3795. The line represents
the minor axis direction. Bottom:
Histogram representing the number of QSOs as a function of
position angle with respect to the minor axis of NGC 3795 within
3 deg of it. 
A constant dependence is excluded within the 7.2$\sigma $ level.}
\label{Fig:histoPA_NGC3795}
\end{figure}

Two of the galaxies show a level of anisotropy higher than
$3\sigma$. The galaxy with highest anisotropy is PGC 36192 [NGC 3795].
See the distribution of QSOs around it in 
Figure~\ref{Fig:histoPA_NGC3795}. The excess $E_{45}$ (excess
of QSOs of this galaxy
in the region that are at position angles less than 45$^\circ $ from
the minor axis over those that have position angles larger
than 45$^\circ $ from it) is +99 (5.2$\sigma $ over zero), 
and for this particular
case $\alpha =2.02\pm 0.28$, an excess of 200\% towards the minor
axis with respect to the major axis.

Among the 71 galaxies for $\theta _{\rm max}=3^\circ$, there are (according to the
SIMBAD database, see Table \ref{Tab:indiv}): 
one LINER, two interacting galaxies, four low surface brightness
galaxies, 15 galaxies in pairs/groups/clusters, and 49 normal galaxies. 
We looked for signs of  correlation of the
galaxy Hubble type with the anisotropy and found nothing. 
There is no correlation of $E_{45}$ with the position angle of the galaxy
(see Table \ref{Tab:indiv}).
There is some trend of higher anisotropy with lower distance (see Table \ref{Tab:indiv}):
only  $5(\pm 5)$\% (1/19) of the galaxies with $d>50$ Mpc have $E_{45}>2\sigma (E_{45})$,
while  for $27(\pm 7)$\% (14/52) of galaxies $d<50$ Mpc 
with $E_{45}>2\sigma (E_{45})$. This may be because 3 deg is
 too great a distance for galaxies with $d>50$ Mpc and the possible effect of
anisotropy is diluted in them. 
There also seems to be a tendency for interacting galaxies to have high anisotropy 
(see Table \ref{Tab:indiv}), but good statistics cannot
be carried out with only two galaxies out of 71. And we have only one Seyfert (LINER),
so neither can we comment on  statistics of this type.
 
\section{Statistical significance of the anisotropy}
\label{.tests}

The spectroscopic survey SDSS is not complete for all redshifts up to the
magnitude $m_g\le 21$. Moreover, we have observed that the completeness
depends on position in the sky because the target selection varies from
region to region. There are, for instance, important gradients of QSO density
with the right ascension.  Moreover, there are selection effects depending  on
the redshifts of the sources; SDSS is not equally capable of detecting sources at
all redshifts, etc. Nonetheless, it is appropriate for our statistical analysis
because that completeness has nothing to do with the orientation of nearby
galaxies; that is, whether there are more or fewer QSOs has nothing to do with the
presence of a galaxy some degrees away and even less with the orientation of the latter.

There will be a global effect in the dispersion from the isotropy due to the
different gradients and systematics, which will be higher with respect to a
perfect statistical survey, so we cannot trust the  statistical error bars
assuming a random distribution.  For instance, the 7.8$\sigma $ level which we
derived to be the significance of non-zero slope of the fit of expression
(\ref{C}) applied to Fig. \ref{Fig:histoPA} would indicate the level of anisotropy
of the Schneider et al.\ (2005) sample, which is both due to the anisotropy in the
real distribution and to the systematic deviations of a homogeneous sample. We
can also calculate this level  of significance through Kolmogorov--Smirnov test
applied to the same data, which also gives a low probability of being compatible
with an angle-independent  distribution: $P_{K-S}=2\times 10^{-8}$ (equivalent to
5.6 $\sigma $ in a Gaussian distribution). Or simply we can count in the same
data the excess in the range 0--45$^\circ $ (13\,025 cases) over the range
45--90$^\circ $ (12151 cases) in Fig. \ref{Fig:histoPA}: $E_{45}=+874$ cases
(Poissonian error $\pm 159$), i.e.\ 5.5$\sigma $. In order to distinguish which
part of the significance of the anisotropy is due to the real distribution, we
have to carry out Monte Carlo simulations.

\subsection{Monte Carlo simulations with random position angle of the galaxies}
\label{.testangle}

\begin{figure}
\begin{center}
\mbox{\epsfig{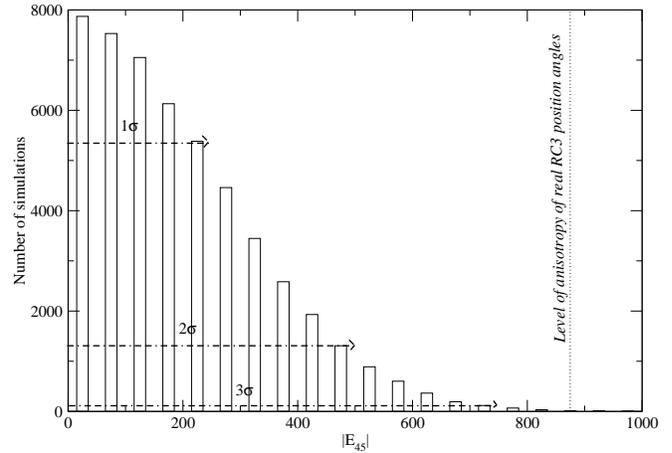}}
\end{center}
\caption{Results of the $|E_{45}|$, excess in the range 0--45$^\circ $
over the range 45--90$^\circ $ (with a total of 25\,176 galaxy--QSO associations), 
of a Monte Carlo simulation with random position angles
for the RC3 galaxies. As shown, the real position angles give a significant
anisotropy of $E_{45}=874$, which is the 3.5$\sigma $ level.}
\label{Fig:histosimul}
\end{figure}

Given the present distribution of QSOs in Schneider et al.\ (2005), we have
randomly generated the position angles of 71 galaxies in the RC3
(the correlation of position angles, if any, are small and insignificant; 
see \S \ref{.randomQSO}), and we have
measured the same numbers $P_{K-S}$ and $E_{45}$. We have carried 50\,000 Monte Carlo
simulations of this type. The distribution of $E_{45}$ for the case without
constraints is shown in Figure~\ref{Fig:histosimul}. This distribution results in a
Gaussian distribution with  $\sigma \approx 250$. Our detection of anisotropy would
be at the  3.5$\sigma $  level.  Table \ref{Tab:montecarlo} presents the results and the statistical
significance for this and other cases with further constraints.

It is remarkable that merely by removing the cases with $z<0.5$ (only 9\% of the cases)
we increase the statistical significance considerably  (to 3.9$\sigma $). 
This is due to the negative
signal of the anisotropy for these removed cases, as shown in Figure
\ref{Fig:ratio_z}. As said, this better signal/noise of the anisotropy  is also observed when
comparing the two cases shown in Figure~\ref{Fig:twod}.

\begin{table*}
\caption{$\sigma $ from the results of Monte Carlo simulations with random 
orientations of the galaxies for the measured values of $P_{K-S}$ and
$|E_{45}|$ in the real distribution.}
\begin{center}
\begin{tabular}{cccccc}
\label{Tab:montecarlo}
Constraint &  N$_{pairs}$ & $P_{K-S}$ & $\sigma (P_{K-S})$ & $E_{45}$ & $\sigma
(|E_{45}|)$  \\
\hline
$\theta _{max}=3^\circ $ & 25176 & $2\times 10^{-8}$  & 3.5 & 874 & 3.5 \\
$\theta _{max}=3^\circ $, $m_g>19.4$, $0.2<\log _{10}(1+z)<0.5$ &  7509 & $5\times 10^{-12}$ & 3.3 & 601 & 3.3 \\
$\theta _{max}=3^\circ $, $m_g>19.4$ & 10204 & $4\times 10^{-10}$ & 3.2 & 644 & 3.4 \\
$\theta _{max}=3^\circ $, $z>0.5$ & 22953 & $8\times 10^{-10}$ & 3.9 & 919 & 3.9  \\
\end{tabular}
\end{center}
\end{table*}

\subsection{The same test rejecting the three farthest outliers}

We repeat the statistical analysis
rejecting the three farthest outliers identified in Table \ref{Tab:indiv} and 
Figure \ref{fig_indiv_gal}. By definition, outliers do not behave like 
the others, either because of the physics or because of certain uncontrolled factors; 
separating outliers might be a test to better understand the properties 
of the general population, and of the outliers themselves if they are physically 
different. We repeat the case only with the constraint $\theta _{\rm max}=3^\circ $.

We now remove three galaxies from Table \ref{Tab:indiv}: PGC 28741,
PGC 36192 and PGC 51752, all of them with the highest positive value of
$E_{45}$, so it is expected that we are going to reduce the signal of the anisotropy.
The number of galaxies is now 68. The number of pairs is 23\,970. $E_{45}=632$, 
$\alpha =0.097\pm 0.018$, $P_{K-S}=4\times 10^{-5}$. The Monte Carlo
simulations on the random orientations of the galaxies
give a significance  $\sigma (P_{K-S})=2.7$, $\sigma (E_{45})=2.6$.
It seems that there is nothing wrong with the previous statistics. The signal
is still there in spite of the removal of the three farthest outliers. It is lower
than 3.5$\sigma $ and is reduced to 2.6 or 2.7$\sigma $, but this is precisely
what would be expected if we remove members that are known to have
the highest anisotropy. There is an average trend for a positive $\alpha $ 
that is not due to the presence of a single  galaxy (or two or three) with some
special circumstances.

\subsection{Tests with subsamples}

We now perform the same calculations with subsamples containing 
90\% of the identified QSO--galaxy pairs. 
If the sample is pure, the anisotropy will decrease slightly owing to the 
smaller number of objects. If outliers have a strong effect, most of the 
simulations will have a slightly higher anisotropy and a few of them will 
have a much lower anisotropy (when the outliers are randomly rejected).

\begin{figure}
\begin{center}
\mbox{\epsfig{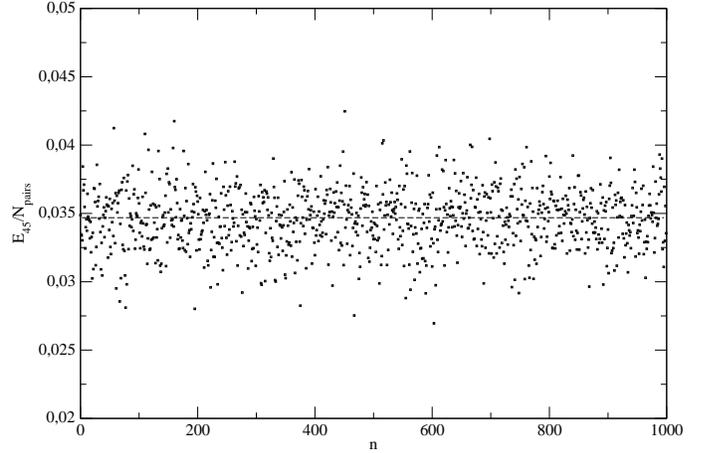}}
\end{center}
\caption{Distribution of values of the relative excess of QSOs with position
angle less than 45 degrees
for 1000 subsamples ($n$ from 1 to 1000), where each subsample
has 90\% randomly selected pairs from the initial 25\,176 in 71 galaxies.
The dashed line represents the value for the original 25\,176 pairs:
$E_{45}/N_{\rm pairs}=0.0347$.}
\label{Fig:e45test}
\end{figure}

For this test with 1000 different subsamples (each subsample
with 90\% randomly selected pairs from the initial 25\,176 in 71 galaxies), 
we calculate the quantity $E_{45}/N_{pairs}$,
which gives a measure of the anisotropy. This time we do not calculate 
the significance for each subsample, because this is very 
time-consuming; in any case, the significance does not need to be calculated for
each different subsample because it is more or less proportional
to $E_{45}/N_{\rm pairs}$.
The results of these 1000 subsamples are shown in Figure 
\ref{Fig:e45test}. The median value of $E_{45}/N_{\rm pairs}$ in these
1000 subsamples is 0.0347, the same value that was obtained
with the original total sample of 25\,176 pairs.
Nothing anomalous is present in this distribution
of values of $E_{45}/N_{\rm pairs}$ so we do not think that a few outliers
are producing the anisotropy but that it is a characteristic of the average
distribution. All 1000 subsamples show a relative excess of QSOs with position
angle less than 45 degrees of $E_{45}/N$ over 0.025.

\subsection{Monte Carlo simulations with random QSO distributions}
\label{.randomQSO}

In the previous subsection, we  showed that, given the actual distribution of
SDSS QSOs, a random distribution of position angles in the RC3 galaxies should not
show the level of anisotropy found; that is, the anisotropy is not due to the
peculiarities of the QSO distribution itself. Now from the position and orientation of
the 71 RC3 galaxies we generate a random Poissonian and homogeneous distribution
of background sources through a Monte Carlo simulation and we shall see whether the
observed level of anisotropy is attained. These should serve to test that the 71
RC3 galaxies have no significant bias responsible for the anisotropy. 
In fact, we have also checked that the distribution of position angles 
in our sample (see Table \ref{Tab:indiv}) is compatible with being randomly 
distributed (according to the Kolmogorov--Smirnov test: a probability of 83\%
for the hypothesis of null correlation). There is no significant correlation
of position angles, although, of course, some small correlation within the noise cannot 
be rejected; in any case, this small correlation would be negligible for 
our statistical purposes.

The results indicate that the probability of obtaining an anisotropy of the measured values
is roughly equal to the probability given by the Kolmogorov test. From 873
simulations, we got 89.9\% with $P_{K-S}>0.1$, 9.5\% with  $0.01<P_{K-S}<0.1$ and
0.5\% with $0.001<P_{K-S}<0.01$; none  with $P_{K-S}<0.001$. This distribution is to be
expected if the probabilities obey the Kolmogorov--Smirnov
distribution. These results indicate that the probability of getting the 
observed value $P_{K-S}=2\times 10^{-8}$
from a random distribution of QSOs is negligible. This means that there is no 
selection effect in the RC3 sample.

\subsection{Tests with other
catalogues than Schneider et al.\ (2005)}

We have also carried out the same test with the spectroscopic galaxies of the 3rd
SDSS release up to $m_g=21$ (333\,314 galaxies), instead of the QSO sample,
and we have found that those galaxies with a difference of velocity with respect
to RC3 galaxies 
higher than 3000 km/s (to avoid their being associated with the
main RC3 galaxy) have an insufficiently significant level of anisotropy.
As always, the RC3 galaxies are spiral and nearly edge-on (inclination greater than 65 degrees).
The result gives $\alpha =-0.029\pm 0.010$ (more galaxies towards the major
than towards the minor axis), $P_{K-S}=2\times 10^{-4}$, and the Monte Carlo
simulations (with random orientation of the RC3-galaxies) show that this has a
probability of 38\% (equivalent to only 0.9$\sigma $). 

If instead of Schneider et al.'s\ (2005) vetted QSO catalogue we use the complete
SDSS 3rd  data release for QSOs (we also put the constraint that the difference in
velocity of the QSO and galaxy must be higher than 3000 km/s),  the anisotropy is
detected with $\alpha =0.111\pm 0.022$, $P_{K-S}=3\times 10^{-6}$, and, according
to Monte Carlo simulations, the anisotropy has a level  of 2.9$\sigma $. The reason
for the lower significance of the anisotropy (with the Schneider et al.\ sample it was
3.5$\sigma $) is that, apart from further spurious contamination, it includes
QSOs with $M_i>-22$, which are mostly  of low redshift, and this has negative
$\alpha $, as shown in Fig. \ref{Fig:ratio_z}, which reduces the average signal.
With the added constraint $z>0.5$, $\alpha =0.154\pm 0.024$, $P_{K-S}=2\times
10^{-9}$ and the Monte Carlo simulations indicate anisotropy at a level of
3.6$\sigma $.

We can also use the newest release, SDSS 4th (DR4, Adelman-McCarthy et
al.\ 2006), which
covers 4783 sq.\ degrees and contains 61\,049 QSOs with $m_g<21.0$. Although the
area covered is only  14\% greater than that covered by the DR3, the total number of RC3 galaxies
 that have circles of radius 3 degrees totally covered is somewhat higher: 127
instead of 71, and the number of QSO--galaxy associations (some QSOs may have more
than one association with a galaxy) is 47\,271 instead 25\,176. This improves the statistics slightly.
However, since the vetted catalogue has not yet been produced and 
will not be produced until the 5th release is delivered (Schneider, priv.\ comm.)
only the analysis of the anisotropy of the complete SDSS 4th data release is
possible, as in the previous paragraph. The result is:  $\alpha =0.073\pm
0.013$, $P_{K-S}=2\times 10^{-5}$, and according to Monte Carlo simulations the
anisotropy has a level of 2.5$\sigma $.  With the extra constraint $z>0.5$, $\alpha
=0.106\pm 0.012$, $P_{K-S}=7\times 10^{-9}$ and the Monte Carlo simulations suggest
 anisotropy at a level 3.3$\sigma $. The average anisotropy is somewhat lower
than with only DR3 sources, possibly the first 71 galaxies have on average a
higher anisotropy; in any case, the differences are roughly within the
expected random fluctuations from one sample to the other.

\begin{figure}
\begin{center}
\vspace{1cm}
\mbox{\epsfig{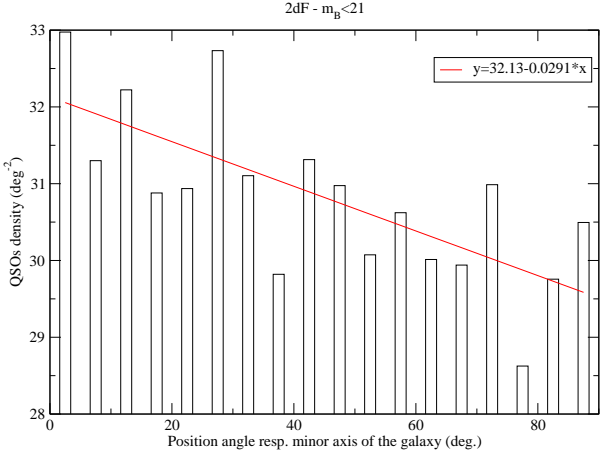}}
\end{center}
\caption{Histogram representing the 2dF QSO density with $m_B<21$
within a circle of radius equal to 3 degrees of 26 RC3 galaxies (more
than 80\% of the circle covered) as a function of position 
angle with respect to the minor axes of these galaxies.}
\label{Fig:histo2df}
\end{figure}

Another search for anisotropy was made using the 2dF QSO spectroscopic
catalogue (Croom et al.\ 2004) with $m_B<21$ (23\,660 QSOs in total). 
This catalogue has the disadvantage of a much lower coverage, 
around 700 square degrees, and its two strips have a width of  5 degrees, which
does not allow complete circles of radius 3 degrees (we consider
the full areas of the two 75 deg $\times 5$ deg).  
We do the analysis with nearly edge-on spiral 
RC3 galaxies that have at least 80\% of the circles
with $\theta _{\rm max}=3^\circ $ covered (26 galaxies) 
and we normalize the QSO counts by
dividing by the covered area. This results in  poorer statistics
than SDSS, but it is illustrative that we also observe the anisotropy,
although with very low significance.
Figure \ref{Fig:histo2df} shows a histogram of densities:
$\alpha =0.089\pm 0.022$, $E_{45}=1.6\pm 0.5$ deg$^{-2}$; a
probability according to Monte Carlo simulations of 51\% (0.7$\sigma $),
which is too insignificant to be considered a detection of
anisotropy but  is compatible with the results from the SDSS survey.
If instead of  80\% minimum coverage, we use only galaxies with  90\% minimum
coverage, we have 13 galaxies, $\alpha =0.109\pm 0.028$, 
$E_{45}=2.1\pm 0.6$ deg$^{-2}$; and a
probability according to Monte Carlo simulations of 22\% (1.2$\sigma $),
still insignificant. For $\theta _{\rm max}=2.5^\circ $, and a coverage
greater than 90\%: 20 galaxies, $\alpha =0.084\pm 0.017$, 
$E_{45}=1.2\pm 0.6$ deg$^{-2}$; and a
probability according to Monte Carlo simulations of 34\% (1.0$\sigma $),
insignificant.

\section{Some considerations towards  measuring the anisotropy}

\subsection{Anisotropy as a function of linear distance}

\begin{figure}
\begin{center}
\mbox{\epsfig{file=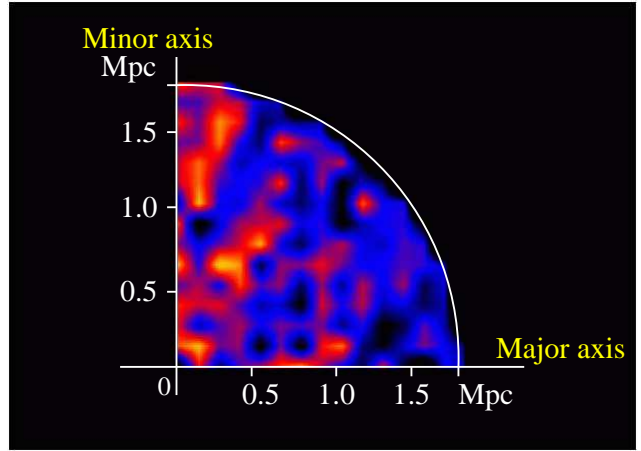,height=6cm}}
\end{center}
\caption{Counts of QSOs (total: 16\,572) as a function of position with
respect to the minor/major axis angles of the corresponding 113 galaxies 
for linear distances lower than 1.8 Mpc (see text). 
Counts were plotted in bins of 0.12 Mpc $\times 0.12$ Mpc
(average around 110 QSOs per bin) and smoothed/interpolated.
The clearest colours indicate higher density. }
\label{Fig:fort93l}
\end{figure}

One possible question about our measure is why we counted the QSOs within a fixed
angular distance (3 degrees) instead of a fixed linear distance. Since we know
the distance of the RC3 galaxies,  we could count the QSOs within a linear
projected distance.  Indeed, we tried to measure the anisotropy in such a way and
we also detected the anisotropy but the statistical significance was not as high
as 3.5$\sigma $ (3.9$\sigma $ for  $z_{\rm QSO}>0.5$). With identical criteria,
using Schneider et al.'s (2005) QSO catalogues but for the maximum linear distance
of 1.8 Mpc (where we got the best values of the anisotropy) instead of 3 degrees
of angular distance, the results are:  113 RC3-galaxies, 16\,572 
galaxy--QSO associations, $\alpha =0.138\pm 0.038$, $P_{K-S}=2\times 10^{-5}$, with a significance
according to Monte Carlo simulations of 2.6$\sigma $. 
The two-dimensional distribution as a function of distance is plotted in
Figure \ref{Fig:fort93l}.
For $z_{\rm QSO}>0.5$, there are
15\,079 galaxy/QSO associations,  $\alpha =0.172\pm 0.042$, $P_{K-S}=2\times
10^{-7}$, with a significance according to Monte Carlo  simulations of 3.1$\sigma $. 

This slightly lower significance of the anisotropy may be due
to the following reasons: i) the selected 
number of very nearby galaxies ($\sim 10$ Mpc) with the linear
distance criterion is lower than the angular distance criterion because
most of them do not follow the constraint of being totally covered
by the SDSS 3rd release survey, a galaxy at $\sim 10$ Mpc would give
a circle with radius $\sim 9$ degrees, which is very unlikely to be totally
covered by the SDSS-3rd release survey (and, as said in \S \ref{.indiv},
there is a slight trend towards higher anisotropy for lower distances); 
ii) the number of QSOs per galaxy is lower for more distant galaxies, because they cover
a lower area. These arguments can explain the
numbers we obtained. 

One might restrict the range of distances to avoid this
selection effect in some measure. For instance, if we take only the RC3 galaxies
within the range of distance $20<d<40$ Mpc (around the median value of 32 Mpc;
i.e.\ $0.005<z_{\rm gal}<0.010$ instead of $z_{\rm gal}<0.050$). We would then have 20
RC3 galaxies, 9457 QSO--galaxy associations , $P_{K-S}=2\times 10^{-7}$, Monte Carlo
significance: 2.8-$\sigma $. For $z_{QSO}>0.5$: 8631 QSO--galaxy associations,
$P_{K-S}=1\times 10^{-9}$, Monte Carlo significance: 3.3$\sigma $. The
signal/noise is a little better with this restriction but still less than the
values obtained with the condition $\theta <3^\circ $. Here, the reason is
possibly the lower number of galaxy--QSO associations used for the statistics.
Whatever  the reason, it seems that the condition $\theta <3^\circ $ gets
higher anisotropy than $linear\ separation<1.8$ Mpc.

\subsection{Counting some QSOs twice or more times}

\label{.twice}

An interesting question is how to select QSO/galaxy pairs.
At present, we have selected all the QSOs  around a given RC3 galaxy;
consequently, if two galaxies are closer together than 6 degrees, they will have some
common QSOs within the circle of radius 3 degrees; some QSOs count twice or more
in the QSO/galaxy association. Since the position and orientation of the
galaxies are random, the different counts with the same QSO should be independent.
Even in the case that they were not independent, the Monte Carlo simulations of
\S \ref{.testangle} would include the possible factors due to it, and the
significance derived therein would contain them. We have estimated the
anisotropies with other criteria in which a QSO counts only once: associating it
with the closest/farthest galaxy in the case that two or more galaxies include it
within their circles, and the observed anisotropy has only slight differences:
$E_{45}/N_{pair}=0.0328/0.0373$ respectively for closest/farthest criterion
instead of 0.0347 (Table \ref{Tab:montecarlo}). This is
explicable perhaps because we would be adding a selection effect in which the
QSOs are on average closer or farther from the galaxies; or by coordinates, in
which a QSO is associated with the first/last galaxy in the list of RC3, which is
ordered by coordinates, and in this case we found more notable differences,
which we attribute to the gradient of QSO density 
in right ascension present in SDSS survey. Summing up,
the most objective criterion is to include all QSOs around a galaxy even if they
are counted twice or more; other criteria can introduce extra bias;
moreover, we have fewer counts and consequently less significance in the detection
of the anisotropy, although whatever   the criteria, the Monte Carlo
simulations will include them and the significance will be correctly calculated.

\subsection{With incomplete circles}
\label{.incomp}

So far, we have used circles that are covered completely
by the SDSS survey (except for 2dF data).
We have performed other measures with circles partially covered,
normalizing the number of counts with the area covered per segment,
and we have observed that: i) it is a trivial result that 
the number of galaxies is higher for lower minimum allowed coverages;
ii) lower minimum coverages in the circles produce higher dispersion and 
lower average anisotropy in our case. When 
the non-covered holes are larger, random effects dominate
[perhaps because we are taking border areas with significant gradients of completeness;
also, the errors in the determination of the areas of the segments might contribute]
and in our case the average random anisotropy is fortuitously negative.
We have perfomed measures only with galaxies of low coverage and we have tested
that they produce a negative $\alpha $, in which Monte Carlo simulations show that
it is not significant. When we mix low-coverage with full-coverage circles we
are wrecking/smoothing the effect observed with the full-coverage circles alone.
 
The best compromise between points i) and
ii) to get the highest significance of the anisotropy was found for the constraint of
98\% as minimum allowed coverage. This gives 135 galaxies (instead of 71 galaxies
with total coverage), 48\,131 QSO-galaxy pairs ,
$\alpha =0.133\pm 0.013$, $E_{45}=0.837$ deg$^{-2}$ whose significance
according to Monte Carlo simulations is 4.4$\sigma $ (a probability of
1 in 90\,000). For $z>0.5$ the significance according to 
Monte Carlo simulations is 4.8$\sigma $
(a probability of 1 in 600\,000). Table \ref{Tab:incomp} gives further values
for other coverages.

\begin{table*}
\caption{Statistics with different minimum allowed coverages of the circles with
$\theta _{\rm max}=3^\circ $.}
\begin{center}
\begin{tabular}{cccccc}
\label{Tab:incomp}
Min.\ cover. & $N_{\rm gal}$ & $N_{\rm pairs}$ & $\alpha $ & $E_{45}$ (deg$^{-2}$) & $\sigma (|E_{45}|)$  \\
\hline
100\% & 71 & 25176 & 0.132$\pm 0.017$ & 0.872 & 3.5 \\
99\% & 117 & 41888 & 0.124$\pm 0.015$ & 0.791 & 4.0 \\
98\% & 135 & 48131 & 0.133$\pm 0.013$ & 0.837 & 4.4 \\
98\% ($z_Q>0.5$) & 135 & 43839 & 0.158$\pm 0.014$ & 0.903 & 4.8 \\
97\% & 143 & 50938 & 0.105$\pm 0.013$ & 0.677 & 3.8  \\
95\% & 162 & 57000 & 0.084$\pm 0.010$ & 0.546 & 2.9 \\
90\% & 188 & 64791 & 0.067$\pm 0.011$ & 0.452 & 2.5 \\
\end{tabular}
\end{center}
\end{table*}

\section{Interpretation}

\begin{figure}
\begin{center}
\mbox{\epsfig{file=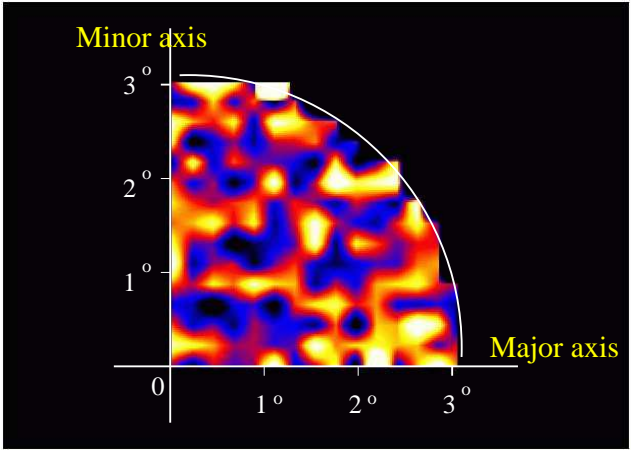,height=6cm}}
\end{center}
\caption{Distribution of the average colour $\langle (g-r)\rangle$ 
as a function of position with
respect to the minor/major axis angles of the corresponding 71 galaxies 
(see text). Average: 0.173. Bins of $0.2^\circ \times 0.2^\circ $,
smoothed/interpolated. The clearest colour indicate reddest colour.}
\label{extin}
\end{figure}

This distribution of objects around galaxies is similar to that found by 
Zaritsky et al.\ (1997) and Azzaro et al.\ (2006)
(the Holmberg effect), i.e.\ a higher density of objects in the minor
axis  direction. However, their distribution was for satellite galaxies associated 
with the main galaxy and our objects are QSOs with very different redshifts with 
respect to the main galaxy. The projected 
scales are also different  $\sim$ 300--500 kpc for
the satellites, and $\sim$ 1--1.5 Mpc in our case. The Holmberg effect has been
explained as a consequence of the preferential capture by the parent galaxy of
satellites orbiting near the equatorial plane. From a purely phenomenological point
of view, the effect discovered here is similar to the Holmberg effect. Although we
have not tried to build a detailed model, we have considered three
possibilities:

$\bullet$ {\bf Extinction}: It might be  that there is some
extinction along the major axis of the galaxies that is high enough to reduce
significantly the number of QSOs observed in that direction. The 
affected range of redshifts ($z>0.5$) contain QSOs in which H$_\alpha $
lines cannot be detected in the optical SDSS survey
(and Ly-$\alpha $ cannot be observed for $z<2.2$),
so lines like H$_\alpha $ (or Ly-$\alpha $) are not used for their identification
as QSOs when $0.5<z<2.2$; instead, other, fainter lines are used, 
and they will be near the limit of detection at $m_g$ around 20.
An excess of extinction along the major axis of $\sim 0.1$ mag 
could produce a selection effect in the SDSS survey
that removes some of the objects to be classified as QSOs
because the signal/noise of some lines becomes
lower than the limit of detection. For example, if we have a QSO
whose equivalent width of the most intense line has got a signal/noise
of 4 without extinction, with extinction this signal/noise would be
lower than 4, so the line of this object would not be catalogued and therefore
this object would not be classified as a QSO.
However, it is difficult to accept that some clouds along the major axis 
associated with a galaxy are as far as 1.5--2.0 Mpc (see Fig. \ref{Fig:fort93l}) 
from it. Intergalactic clouds might reside at 1.5--2 Mpc from a galaxy
but there is no reason in principle to think they are in the plane
of the galactic disc rather than in a random distribution. A random
distribution of intergalactic clouds would not produce anisotropy.
Moreover, we have
computed the polar and radial distribution of $g-r$ colours of our sample of 
QSOs and we find no appreciable differences between the mean colour of objects at angles 
$\le \ge \pm 45^\circ$ from the minor axis of the galaxy (Figure
\ref{extin} presents the two-dimensional distribution of such colours;
the average colour as a function of the position angle from the minor axis is
$\langle (g-r)\rangle =0.1727\pm 0.0022 + 
1.9\pm 4.3\times 10^{-5}PA({\rm deg})$), so
unless a grey-dust extinction is present, this explanation does not
seem appropriate.

$\bullet$  Gravitational lensing: As said in the introduction, 
previous detection 
of a correlation between high redshift QSOs and low redshift galaxies has been
tentatively attributed to effects of gravitational lensing in the halo of
foreground galaxies. This solution does not work in general
(Zhu et al.\ 1997; Tang \& Zhang 2005), and
it  also has important problems in explaining the phenomenon of this paper:

\begin{itemize}

\item  The dependence of the anisotropy on the redshift
of the QSOs (excess over $z>0.5$ and defect over $z<0.5$)
cannot be explained in terms of the distance of the source and the lens.
All gravitational lens effects, whatever  the model,
have an Einstein radius $\theta _E$ such that $\theta _E^n$ is
proportional to $\frac{D_{ls}}{D_s}$ (e.g.\ $n=2$ for microlensing,
$n=1$ for singular isothermal sphere approximation), 
with $D_{ls}$ and $D_s$ respectively
the distance lens--source, source--observer.
$\frac{D_{ls}}{D_s}$ is nearly one for all QSOs at $z>0.08$
(assuming a distance of the galaxy of 32 Mpc, $\frac{D_{ls}}{D_s}>0.9$). We
cannot understand with this hypothesis the absence of anisotropy
for most QSOs with $z<0.5$. The dependence on the magnitude
shown in Fig.\ \ref{Fig:counts} is also difficult to understand.
we cannot understand why there is no lensing effect
for the brightest QSOs.

\item The linear scales of the effect discovered here seem too
large for the size of the halos as estimated in $N$-body CDM or $\Lambda$CDM
simulations. For instance, typical virial radii are $\sim$ 100--300 kpc in these
models. Although Prada et al.\ (2005) have demonstrated that actual halos extends
at least up to 2--3 $R_{\rm vir}$, the density at this distances is so low that it seems
difficult to produce significant lensing effects. A point to note is
that such simulations predict elongated halos (Allgood et al.\ 2006; 
Mandelbaum et al.\ 2006), whose major axes in the inner
parts tend to be orthogonal to the galactic discs, while in the outer parts the
halos tend to be randomly orientated. The anisotropy found in this work would 
 need elongated structures perpendicular to the galactic discs and
extending out to $\sim 1.5$ Mpc (see Fig. \ref{Fig:fort93l}).

\item  The amount of mass necessary is huge.
Assuming as a first approximation that the increase in counts in SDSS is due 
to an increase in the magnitudes of the background sources (this is not
exactly true, because the magnitudes of an object refer to the
average flux of the object in a passband, while the SDSS detection
criteria depend on the signal/noise of some lines, but the order of magnitude
should not be very different), the average enhancement is
$\langle q\rangle \approx 1.07$ (an increase of 7\% in the counts) 
in the counts over $\approx 14$ deg$^2$ (the region within $\theta <3^\circ $
and polar angle within 45$^\circ $ from the minor axis).
The magnification factor $\mu $ due to gravitational lensing 
[see, for instance, eq. (1) of Zhu et al.\ 1997] is related to $q$ by
means of
\begin{equation}
q=\frac{N(m_b+2.5\log _{10}\mu )}{N(m_b)}\frac{1}{\mu }
,\end{equation}
where $N$ is the cumulative QSO counts up to magnitude $m_b$.
With the Boyle et al.\ (2000) counts [$N=1981-214.2m_b+5.792m_b^2$ (L\'opez-Corredoira
\& Guti\'errez 2004, appendix A);
in the real distribution of QSOs, not in our sample which is incomplete],
$\langle \mu \rangle \approx 1.28$ for $m_b=20.0$ on average over 
$\approx 14$ deg$^2$. For $m_b=19.5$, it is $\langle \mu \rangle \approx 1.08$
and for $m_b=20.5$ there is no value of $\mu $ that gives an enhancement as high as
$q=1.07$. Let us consider the average value of $m_b\approx 20.0$: $\langle \mu \rangle \approx 1.28$. 
With a single isothermal sphere approximation $\mu \approx
\frac{\theta }{\theta -\theta _E}$, where $\theta _E$ is the Einstein
radius; whose average over a circle
of radius $\theta _{\rm max}>>\theta _E$ 
is $\langle \mu \rangle \approx 1+ \frac{2\theta
_E}{\theta _{\rm max}}$. Therefore, in our case it would be
$\theta _E\approx 1100$ arc seconds, which would require an excess
of mass in the minor axis region with respect to the major axis region of a 
whole rich cluster of galaxies (for the typical
galaxy distance of 32 Mpc) in a single spiral galaxy (Wu 1996). 
Even if the mass is distributed over many lenses, the total necessary
mass is of the same order because $\langle (\mu -1)\rangle \propto
\sum _i \theta _{E,i}\propto \sum _i M_i=M_{\rm total}$, 
although the numbers can change depending on the size of the lenses.
With a point-like microlensing approach (Paczy\'nski 1986; Wu 1996), 
$\mu \approx \frac{u^2+2}{u\sqrt{u^2+4}}$, $u=\theta /\theta _E$, 
whose average over a circle of radius $\theta _{max}>>\theta _E$ 
is $\langle \mu \rangle \approx 1+ \frac{2\theta
_E^2}{\theta _{max}^2}$, it results in $\theta _E=2900$ 
arc seconds. In the case of microlensing, $\theta _E^2$ is proportional
to the mass ($\theta _E^2=\frac{4G\ M}{dc^2}$ for distances
of the QSOs much greater than the distance of the galaxy; 
Paczy\'nski 1986; Wu 1996; Tang \& Zhang 2005),
$\langle (\mu -1)\rangle \propto
\sum _i \theta _{E,i}^2\propto \sum _i M_i=M_{\rm total}$, 
so again it does not matter whether the mass is distributed in 
 one or  many sources: the
average magnification will be the same. In our case, we
would need a mass $M\sim 4\times 10^{16}$ $M_\odot $ 
for $d=32$ Mpc, an impossible value.

\end{itemize}

$\bullet$ {\bf Non-cosmological redshifts}: 
At least some QSOs are associated with nearby
parent galaxies. They would have absolute magnitudes between $-$10 and $-$14 in $g$. 
In this unorthodox scenario, QSOs might be ejected with velocity
enough to become gravitationally unbounded to the galaxy.
Arp (1998a, Fig. 3-27) proposes ejection along the minor axis, although
out to distances of 0.5 Mpc. 
If we were to assume this
hypothesis to be correct, why do we find this effect at distances up to 2--3$^\circ $? 
Maybe because our sample have relatively faint QSOs, and QSOs with small angular
separation from a galaxy  have  higher luminosities on average (Dravskikh \&
Dravskikh 1996) that correspond to lower apparent magnitudes. 
Possibly the same effect of anisotropy would be
observed for bright QSOs on scales less than 1$^\circ $. 
Bright QSO excess might not be detected because the number of such QSOs in
SDSS is very low for their excess to be detected statistically.
Why do the QSOs with redshift
$z>0.5$ preferably  show this anisotropy and not other redshifts? If Arp's 
hypothesis of ejection were right, maybe QSOs with lower 
redshifts are further away or turned into galaxies, but
this places the maximum distance even beyond $\approx 1.5$ Mpc, so the
problem with distance in Arp's hypothesis still remains.
That interacting galaxies appear among the galaxies with the highest anisotropy in 
the distribution might point some relation between both phenomena. 
Therefore, Arp's hypothesis qualitatively  predicts the observed anisotropy but it
fails in the distance estimation of the QSOs from the galaxies by a factor of 
$\sim 3$.

Could the detected anisotropy be a statistical fluctuation?
It might be, but it is a very low probability one: for instance
3.5$\sigma $ means a probability $5\times 10^{-4}$ 
(with no preselection of QSOs and galaxies except the angle
$\theta _{\rm max}=3^\circ $). 
Moreover, Fig.\ \ref{Fig:ratio_z} shows a clear dependence on redshift which
is not expected for a random sample of QSOs (there is a clear
non-random trend of anisotropy depending on $z$), so the significance
increases to 3.9$\sigma $ only by removing the QSOs with $z<0.5$ (9\%
of the total). Is this possibly due to chance? We think that this is not very
likely, but that is precisely the nub of the question.
Here, we deliver this new challenge in the long-running puzzle of this old 
topic: the relationship between high redshift QSOs and nearby galaxies. 
In any case, since this is the first time that this anisotropy has been observed
and the possibility of a chance fluctuation, although unlikely, is not
totally discarded, we prefer at present to be prudent and just say 
that it is a ``tentative'' detection that should be corroborated by 
other groups before any extraordinary claims are made.

{\bf Acknowledgments:}
Thanks are given to the anonymous referee,
Halton C. Arp (MPIA, Garching, Germany),
Ruben J. D\'\i az (C\'ordoba, Argentina)
and D. P. Schneider (Pennsylvania, USA) for
useful comments, and T. J. Mahoney (IAC, Tenerife, Spain) for proof-reading
of the text. We acknowledge use of RC3 catalogue its 
authors. Funding for the creation and distribution of the SDSS Archive has  been
provided by the Alfred P. Sloan Foundation, the Participating  Institutions, the
National Aeronautics and Space Administration, the  National Science Foundation,
the U.S. Department of Energy, the Japanese  Monbukagakusho, and the Max Planck
Society. The SDSS Web site is  http://www.sdss.org/. The SDSS is managed by the
Astrophysical Research  Consortium (ARC) for the Participating Institutions. The
Participating  Institutions are The University of Chicago, Fermilab, the
Institute for  Advanced Study, the Japan Participation Group, The Johns Hopkins
University,  the Korean Scientist Group, Los Alamos National Laboratory,  the
Max-Planck-Institute for Astronomy (MPIA), the Max-Planck-Institute for 
Astrophysics (MPA), New Mexico State University, University of Pittsburgh, 
University of Portsmouth, Princeton University, the United States Naval 
Observatory, and the University of Washington.
The 2dF QSO Redshift Survey (2QZ) was compiled by the 2QZ survey team from observations 
made with the 2-degree Field on the Anglo-Australian Telescope.
The authors were supported by the {\it Ram\'on y
Cajal} Programme of the Spanish Science Ministry.


\begin{thebibliography}{99}

\bibitem{} Abazajian, K., Adelman-McCarthy, J. K., Ag\"ueros, M. A.,
et al. 2005, AJ 129, 1755

\bibitem{} Adelman-McCarthy, J. K., Ag\"ueros, M. A., Allam, S. S.,
et al., 2006, ApJS 162, 38

\bibitem{} Allgood, B., Flores, R. A., Primack, J. R., Kravtsov, A. V., Wechsler, R.
H., Faltenbacher, A., \& Bullock, J. S. 2006, MNRAS 367, 1781

\bibitem{} Arp, H. C., 1966, Science, 151, 1214

\bibitem{Arp67} Arp, H. C. 1967, ApJ, 148, 321

\bibitem{} Arp, H. C., 1998a, Seeing Red, Apeiron, Montreal

\bibitem{Arp98b} Arp, H. C. 1998b, ApJ, 496, 661

\bibitem{Arp99} Arp, H. C. 1999a, Active Galactic Nuclei and Related 
Phenomena, Y. Terzian, E. Khachikian, and D. Weedman (Eds.), Astronomical
Society of the Pacific, S. Francisco, 347.

\bibitem{} Arp, H. C., 1999b, A\&A 341, L5

\bibitem{} Arp, H. C., \& Hazard C. 1980, ApJ 240, 726

\bibitem{} Arp, H. C., \& Russell, D., 2001, ApJ 549, 802

\bibitem{} Azzaro, M., Patiri, S. G., Prada, F., \& Zentner, A. R.
2006, astro-ph/0607139

\bibitem{Ben01} Ben\'\i tez, N., Sanz, J. L., \& Mart\'\i nez-Gonz\'alez, E.
2001, MNRAS, 320, 241

\bibitem{} Boyle, B. J., Shanks, T., Croom, S. M., et al., 2000,
MNRAS, 317, 1014

\bibitem{} Burbidge, E. M., Burbidge, G. R., Solomon, P. M., \& 
Strittmatter, P., 1971, ApJ, 170, 223

\bibitem{Bur97b} Burbidge, G. R. 1999, in: Cosmological Parameters and 
the Evolution of the Universe, K. Sato (Ed.), Kluwer, Dordrecht, p.\ 286

\bibitem{} Burbidge, G. R., 2001, PASP 113, 899

\bibitem{Chu84} Chu, Y., Zhu, X., Burbidge, G., \& Hewitt, A. 
1984, A\&A, 138, 408

\bibitem{} Croom S. M., Smith R. J., Boyle B. J., Shanks T., Miller L., Outram P. J., 
\& Loaring N. S., 2004, MNRAS, 349, 1397

\bibitem{} de Vaucouleurs, G., de Vaucouleurs, A., Corwin, H. G. 
Buta, R. J., Paturel, G., \& Fouqu\'e, P. 1991, Third Reference
Catalog of Bright Galaxies, Springer, New York (RC3) 

\bibitem{Dra96} Dravskikh, A. F., \& Dravskikh, Z. V. 1996, Astron Zh., 73, 
19. Translated into english in: 1996, Astron. Rep., 40, 13.

\bibitem{Gaz03} Gazta\~naga, E. 2003, ApJ, 589, 82

\bibitem{} Guimaraes, A. C. C. 2005, astro-ph/0510719

\bibitem{Kov89} Kovner, I. 1989, ApJ, 341, L1

\bibitem{} L\'opez-Corredoira, M., Guti\'errez, C. M. 2004, A\&A, 421, 407

\bibitem{} Mandelbaum, R., Hirata, C. M., Broderick, T., Seljak, U., \& Brinkmann, J.
2006, MNRAS, 370, 1008

\bibitem{Nar89} Narlikar, J. V., \& Das, P. K., 1980, 
ApJ, 240, 401

\bibitem{} Nollenberg, J. G., \& Williams, L. R., 2005, ApJ 634, 793

\bibitem{} Paczy\'nski, B., 1986, ApJ, 304, 1

\bibitem{} Prada, F., Klypin, A. A., Simonneau, E., Betancort-Rijo, J., Patiri, S.,
Gottlober, S., \& S\'anchez-Conde, M. A. 2006, ApJ, 645, 1001

\bibitem{} Richards, G. T, Nichol, R. C., Gray, A. G. et al. 2004,
ApJS 155, 257

\bibitem{} Schneider, D. P., Hall, P. B., Richards, G. T., et al.,
2005, AJ, 130, 367

\bibitem{} Scranton, R., M\'enard, B., Richards, G. T., et al. 2005, 
ApJ 633, 589

\bibitem{} Tang, S. M., \& Zhang, S. N., 2005, Chin. J. Astron. Astrophys.,
5, 147

\bibitem{} Wu, X.-P. 1996, Fund. Cosmic Phys., 17, 1

\bibitem{} Zaritsky, D., Smith, R., Frenk, C. S., \& White, S. D. M.
1997, ApJ, 478, L53

\bibitem{} Zhu, Z.-H., Wu, X.-P., \& Fang, L.-Z. 1997, ApJ, 490, 31

\end{thebibliography}
\end{document}